\documentclass{ptephy_v1}
\usepackage{cancel}

\begin{document}

\title{Speeding Up Dark Matter With Solar Neutrinos}

\author{Yue Zhang}
\affil{Department of Physics, 
Carleton University, Ottawa, K1S 5B6 Canada \email{yzhang@physics.carleton.ca}}

\begin{abstract}%
We present a novel mechanism of using solar neutrinos to speed up dark matter, inspired by the fact that neutrinos are the most energetic particles from the Sun with a well-understood spectrum. In a neutrino portal dark sector model, we show that dark matter with sub-GeV mass could be accelerated by the $pp$ neutrinos to velocities well above $10^{-3}c$ and capable of depositing large enough energy at direct detection experiments. A crucial ingredient of this mechanism is the dissociation of stable dark matter bound states that exist in Nature. The resulting dark matter velocity distribution bears a strong resemblance in shape to the solar neutrino spectrum. As an application, we derive a leading limit on light dark matter interaction by reinterpreting a recent PICO experiment result.
\end{abstract}

\maketitle

\section{Introduction}
Direct detection is an important approach to unveil the nature of dark matter in the Universe. It is widely assumed~\cite{Goodman:1984dc} that most of dark matter particles in our galaxy move non-relativistically, thus the maximal energy transfer per dark matter scattering is limited by its velocity distribution, typically not more than hundreds of keV. Experimentally, in order to hunt signals with such a small energy deposit, tremendous amount of effort has been made to build low-noise detectors deep underground, setting strong limits on dark matter scattering cross sections~\cite{Schumann:2019eaa,Liu:2017drf}. These limits, however, weaken significantly for dark matter mass below a few GeV, where the available energies due to the scattering fall short of the energy threshold of traditional dark matter detectors. To probe sub-GeV dark matter candidates, one either has to devise new detectors with lower energy thresholds, or consider possibilities where dark matter is made to travel faster~\cite{Bringmann:2018cvk,Ema:2018bih,Cappiello:2018hsu,Wang:2019jtk,Kouvaris:2015nsa,An:2017ojc,Agashe:2014yua,deNiverville:2011it,Izaguirre:2013uxa}, or novel ways of detection beyond the elastic scattering picture~\cite{Grossman:2017qzw,Davoudiasl:2011fj,Dror:2019onn}.

In this article, we present a new mechanism of speeding up (a fraction of) dark matter particles and explore its phenomenological consequences. 
There is room for this possibility thanks to our ignorance of the precise local dark matter velocity distribution~\cite{McCabe:2010zh,Fox:2010bz,Green:2011bv,Frandsen:2011gi}.
A simple way to energize dark matter is through its interaction with high-energy cosmic protons~\cite{Bringmann:2018cvk,Ema:2018bih,Cappiello:2018hsu,Wang:2019jtk}. This is not a local effect and relies on assumptions of cosmic ray distributions further away from the solar system. In contrast, given that the Sun is a powerful energy source nearby which we understand well, it is attractive to consider its impact on the dark matter velocities. 
The dark-matter-electron interaction has been considered to transfer the solar heat to dark matter~\cite{Kouvaris:2015nsa,An:2017ojc} which, however, is found only effective for very light dark matter with mass close to MeV due to the limited solar temperature. Here, it is worth remarking that the most energetic particles from the Sun are not electrons or photons, but neutrinos, which are directly produced by nuclear reactions and mostly escape without thermalization. The typical solar neutrino energies ($\sim$ MeV scale) are orders of magnitude higher than the solar temperature ($\sim$ keV scale). Thus, solar neutrinos are capable of speeding dark matter up to higher velocities. 
Moreover, the solar neutrinos also have a much higher flux than that of the diffuse cosmic rays, and their energy spectrum is well understood in standard solar models~\cite{Bahcall:1987jc} and verified experimentally~\cite{Bellerive:2003rj}.
Motivated by these observations, we investigate the impact of dark-matter-neutrino interaction on the local dark matter velocity distribution and direct detection experiments.

\section{The Kinematics}
Consider dark matter mass lying between tens of MeV and GeV scale, the solar neutrino energy satisfies the hierarchy, $m_\chi v \ll E_\nu \ll m_\chi$, where $v\sim 10^{-3}c$ is the virialized halo dark matter velocity. In this case, the solar-neutrino-dark-matter elastic scattering, 
\begin{equation}\label{vanillaprocess}
\nu + \chi \to \nu + \chi \ ,
\end{equation}
is approximately a fixed-target collision, and the velocity of final state dark matter $\chi$ is given by
\begin{equation}\label{normalv}
v_\chi \simeq \frac{2 E_\nu}{m_\chi} \cos\theta \ ,
\end{equation}
where $\theta$ is the relative angle between the incoming $\nu$ and out going $\chi$ and for simplicity we ignored the halo velocity. 
When $\chi$ travels to a dark matter detector and strikes on a nucleus target, the energy transfer is at most
\begin{equation}\label{eq:Ereco}
E_r \sim \frac{\left(m_\chi v_\chi \right)^2}{2 m_A} \lesssim \frac{2 E_\nu^2}{m_A} \ ,
\end{equation}
where $m_A$ is the target nucleus mass, ranging from 10--100\,GeV. 
Note the dark matter mass dependence drops out.
In order for this recoil energy to surpass the detector's energy threshold, typically of order keV, we would need
$E_\nu \gtrsim 10\,{\rm MeV}$.
From the sun, such an energy is only accessible via the $^8B$ and {\sl hep} neutrinos, which suffer from a much lower flux compared to their low energy counterpart~\cite{Strigari:2009bq}.

This limitation and our ignorance about the nature of dark matter inspire us to go beyond the process in Eq.~(\ref{vanillaprocess}) but instead consider the following,
\begin{equation}
\nu + B \to \chi + \eta \ ,
\end{equation}
where $B$ is a composite state of dark matter, which in the simplest case is made of two $\chi$ particles.
It dissociates after scattering with an energetic neutrino. Because neutrino carries spin, in the final state $\chi$ and $\eta$ must be different particles. Hereafter we assume $\chi$ is a fermion and $\eta$ is a scalar. It is straightforward to work out the kinematics in this case. Assuming $m_\eta \simeq m_\chi$ and $m_B \simeq 2 m_\chi$, the final state dark matter could speed up to
\begin{equation}\label{eq:vafter}
v_\chi \simeq \sqrt\frac{E_\nu}{m_\chi} + \frac{E_\nu}{2m_\chi} \cos\theta \ .
\end{equation}
Because $E_\nu \ll m_\chi$, the first term dominates and has no $\theta$ dependence. This velocity is parametrically larger than that in Eq.~(\ref{normalv}).
In the kinematic region where binding energy $E_b$, and/or the $\chi-\eta$ mass difference $\delta m$ is non-negligible, the above formula is modified to $v_\chi \simeq \sqrt{(E_\nu - E_b - \delta m)/m_\chi}$.
The corresponding recoil energy in the event of dark-matter-nucleus scattering is
\begin{equation}
E_{r} \simeq \frac{m_\chi E_\nu}{2m_A} \ .
\end{equation}
It is also parametrically larger than Eq.~(\ref{eq:Ereco}). The key reason behind is that the  whole solar neutrino energy gets absorbed in this dissociation process.

Through this exercise, we discover a novel way of speeding up sub-GeV dark matter with solar neutrinos which could lead to new prospects for dark matter direct detection experiments. The dark matter particle originally residing in a cosmologically stable bound state gets liberated by a solar neutrino, inheriting an order one fraction of its energy. 
Next, we show that these ingredients can be accommodated in a simple dark sector model.

\section{A Viable Model}
We consider a dark sector with fermionic dark matter $\chi$ and a light scalar dark force $\phi$,
with the following
Yukawa interaction
\begin{equation}\label{Lyd}
\mathcal{L}_{\text{dark-Yukawa}} = y_D \bar\chi \chi \phi \ .
\end{equation}
Within this simple dark sector, it has been shown~\cite{Wise:2014jva,Wise:2014ola} possible of accommodating 
a wide variety of dark matter bound states and offering rich physics cosmologically and experimentally. In particular, it was observed 
that the exchange $\phi$ among dark matter particles leads to an attractive force regardless 
of $\chi\chi$ or $\chi\bar\chi$ system. Bound states occur when the mass of $\phi$ is smaller 
than the Bohr radius, $\sim \alpha_D m_\chi$ where $\alpha_D = y_D^2/(4\pi)$. In the case where $\chi$ 
is an asymmetric dark matter~\cite{Nussinov:1985xr,Kaplan:2009ag} where its stability is due to an approximated $U(1)_D$ global symmetry, 
the $\chi\chi$ and even $n\chi\ (n>2)$ bound states are cosmologically stable. 
These states could be produced in the early universe and comprise part of the dark matter relic density~\cite{Wise:2014jva,Gresham:2017zqi,Gresham:2017cvl,Gresham:2018anj}. Alternatively, even without an asymmetry, the $\chi\bar\chi$ bound states could also be stable due to a conserved $C$-parity, as pointed out in~\cite{An:2016kie}.

\begin{figure}[t]
\centerline{\includegraphics[width=0.5\textwidth]{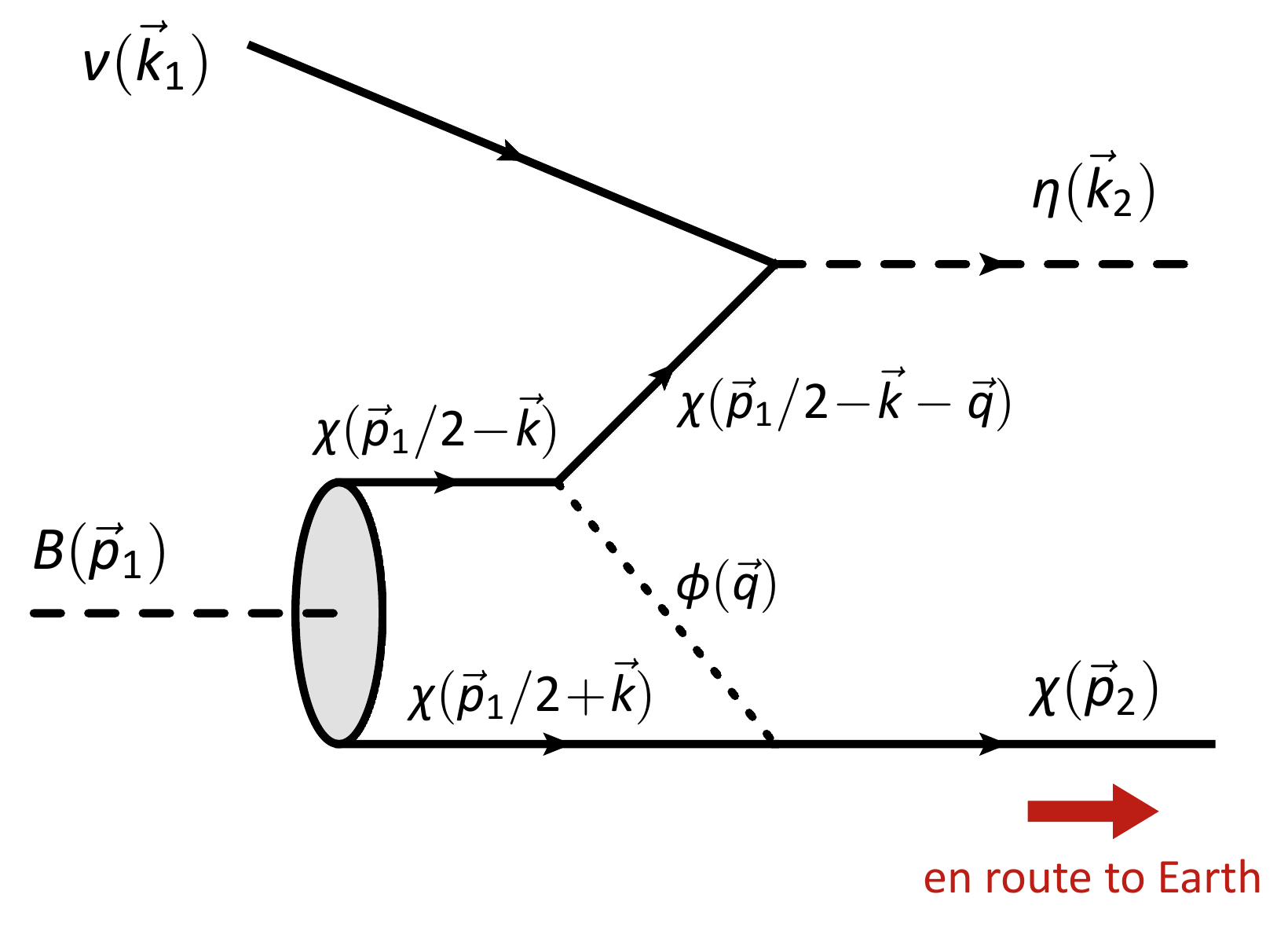}}
\caption{Feynman diagram for solar neutrino to dissociate a dark matter bound state ($B$) and speed up the final state particles, well above the typical halo velocities.}
\label{fig:FD}
\end{figure}

In this work, we will focus on the two-body $\chi\chi$ bound state, assuming it comprises an order one 
fraction of today's dark matter abundance. We restrict our study to the ground state with zero orbital angular momentum. 
Fermion spin statistics then implies that the total spin of the ground state must be zero, {\it i.e.},
$\chi\chi$ must form a pseudoscalar bound state. We call such a composite state $B$ hereafter, which can
be constructed from elementary $\chi$ states, 
\begin{equation}
\left| B (\vec{p}) \right\rangle = \sqrt\frac{1}{2} \sqrt\frac{1}{8m_\chi^3} \sum_{s,s'} \int \frac{d^3 \vec{k}}{(2\pi)^3} \widetilde \Psi (\vec{k}) \left[\bar u_s (\vec{k}) \gamma_5 v_{s'}(-\vec{k}) \rule{0mm}{4mm}\right] \left| \vec{k}, s \right\rangle \otimes \left| -\vec{k}, s' \right\rangle  \ ,
\end{equation}
where $| \vec{k}, s \rangle$ and $| -\vec{k}, s' \rangle$ are two plane-wave $\chi$ particle states, and $\widetilde \Psi (\vec{k})$ is the
bound state wavefunction in the momentum space. $\sqrt{1/2}$ is a symmetry factor accounting for
identical particles.


On top of the above dark sector setup, we introduce a neutrino portal interaction for $\chi$, of the form
\begin{equation}\label{EFO}
\mathcal{L}_{\nu\text{-portal}} = \frac{1}{\Lambda} (LH) (\chi \eta) + {\rm h.c.} \ ,
\end{equation}
where $\eta$ is a complex scalar and heavier partner of $\chi$. 
The conservation of global $U(1)_D$ symmetry (for $\chi$ to be asymmetric dark matter) naturally introduces the $\eta$ field which carries opposite charge to $\chi$. 
The neutrino portal has been widely acknowledged as one of the three leading portals to the dark sector, accompanying the photon and Higgs portal dark sector theories. 
Such an operator and its UV completion have been considered in a number of dark sector contexts models~\cite{Orlofsky:2021mmy,Berryman:2017twh,Batell:2017cmf,Bertoni:2014mva,Cherry:2014xra,Falkowski:2009yz,Berezhiani:1995am}.
Here we go beyond the minimal setup by introducing the dark Yukawa interaction Eq.~(\ref{Lyd}).
In order to support the $U(1)_D$ symmetry for $\chi$ to be stable and an asymmetric dark matter, $\eta$ and $\phi$ cannot be the same particle.

The key process we consider is depicted in Fig.~\ref{fig:FD}, where a solar neutrino $\nu$ strikes on a $B$ bound state, gets absorbed
and dissociate it into two unbounded $\chi$ and $\eta$ particles.
The corresponding scattering amplitude takes the form
\begin{equation}
\begin{split}
&\mathcal{A}_{\nu+ B\to \chi + \eta} = \frac{(v/\Lambda)y_D^2}{(q^2 - m_\phi^2)} \bar u_\chi(p_2) \Pi \frac{1}{\frac{\cancel{p}_1}{2} - \cancel{k} - \cancel{q} - m_\chi} v_\nu(k_1) \ ,  \\
&\Pi = \frac{|\Psi(0)|}{\sqrt{16 m_\chi^3}} \left( \frac{\cancel{p}_1}{2} + \cancel{k} + m_\chi \right) \gamma_5 \left( \frac{\cancel{p}_1}{2} - \cancel{k} - m_\chi \right) \ ,
\end{split}
\end{equation}
where $v=246\,$GeV is the Higgs vacuum expectation value, $\Psi(0)$ is the wavefunction at space origin, and the relevant momenta in lab frame are $p_1 \simeq [2m_\chi, \vec{0}]$, $p_2 \simeq [m_\chi + |\vec{p}_2|^2/(2m_\chi), \vec{p}_2]$, $k_1 = [ E_\nu, E_\nu \hat z]$.
From Eq.~(\ref{eq:vafter}) we derive $|\vec{p}_2| \simeq \sqrt{m_\chi E_\nu}$.
In the limit where the neutrino energy $E_\nu$ is much higher than the binding energy $\approx \alpha_D^2 m_\chi/4$, the momentum transfer $q \simeq [ |\vec{p}_2|^2/(2m_\chi), \vec{p}_2]$ will be much larger than the Bohr momentum $k$.
The above amplitude can be further simplified by imposing momentum conservation $p_2 = p_1/2+k+q$. Keeping only the leading terms, the cross section is
\begin{equation}\label{eq:crosssection}
\sigma_{\nu+ B\to \chi + \eta} \simeq \frac{\pi (v/\Lambda)^2  \alpha_D^2|\Psi(0)|^2}{4 \sqrt{m_\chi E_\nu} (m_\chi E_\nu + m_\phi^2)^2} \ ,
\end{equation}
where it is assumed $m_\eta \simeq m_\chi$ for simplicity.
In Coulomb limit ($m_\phi^2 \ll m_\chi E_\nu$), it further simplifies to 
\begin{equation}\label{eq:crosssectionCoulomb}
\sigma_{\nu+ B\to \chi + \eta} \simeq \frac{(v/\Lambda)^2 \alpha_D^5 m_\chi^{1/2} }{32 E_\nu^{5/2}} \ .
\end{equation}
In this derivation we have assumed the momentum transfer $q$ to be much larger than the bound state Bohr momentum $k$. For small enough $E_\nu$, the denominator will be regulated by the non-zero Bohr momentum in the dominator. The phase space shuts off when the neutrino is no longer energetic enough to dissociate the bound state.

The purpose of this calculation is to show that the process of dark matter bound state dissociation with solar neutrinos could have a sizable cross section. Indeed, for a set of sample parameters $\alpha_D = 0.05$, $v/\Lambda = 0.7$, $m_\chi=0.1\,$GeV and $E_\nu = 1\,$MeV, we find $\sigma_{\nu+ B\to \chi + \eta} \simeq 0.6\times 10^{-28}\,{\rm cm}^2$.

One may also question the fate of the $\eta$ particle. In this calculation we assumed it to be degenerate with $\chi$, which is thus stable. If, on the other hand, $\eta$ is heavier than $\chi$, it could decay into a $\chi$ plus a neutrino. The resulting $\chi$ particle is also boosted and will contribute to the dark matter flux derived below.

\section{Accelerated Dark Matter Flux}
So far we have derived the kinematics and dynamics of a new process for solar neutrinos to speed up dark matter by dissociating bound states. 
The next step is to calculate the resulting flux of dark matter $\chi$. To begin with, we need a distribution of dark matter bound states in the vicinity of the Sun.
In this work, we assume that the bound state $B$ comprises an order one fraction of the total dark matter relic abundance, which is shown to be a feasible scenario~\cite{Wise:2014jva}, and in turn an order one fraction of the local density of dark matter in the solar system.
We do not consider the capture of $B$'s by the Sun because for sub-GeV masses the evaporation effect is strong~\cite{Griest:1986yu,Gould:1987ju}.
The velocity gain of dark matter via evaporation is much smaller than what we consider, Eq.~(\ref{eq:vafter}).

\begin{figure}[t]
\centerline{\includegraphics[width=0.618\textwidth]{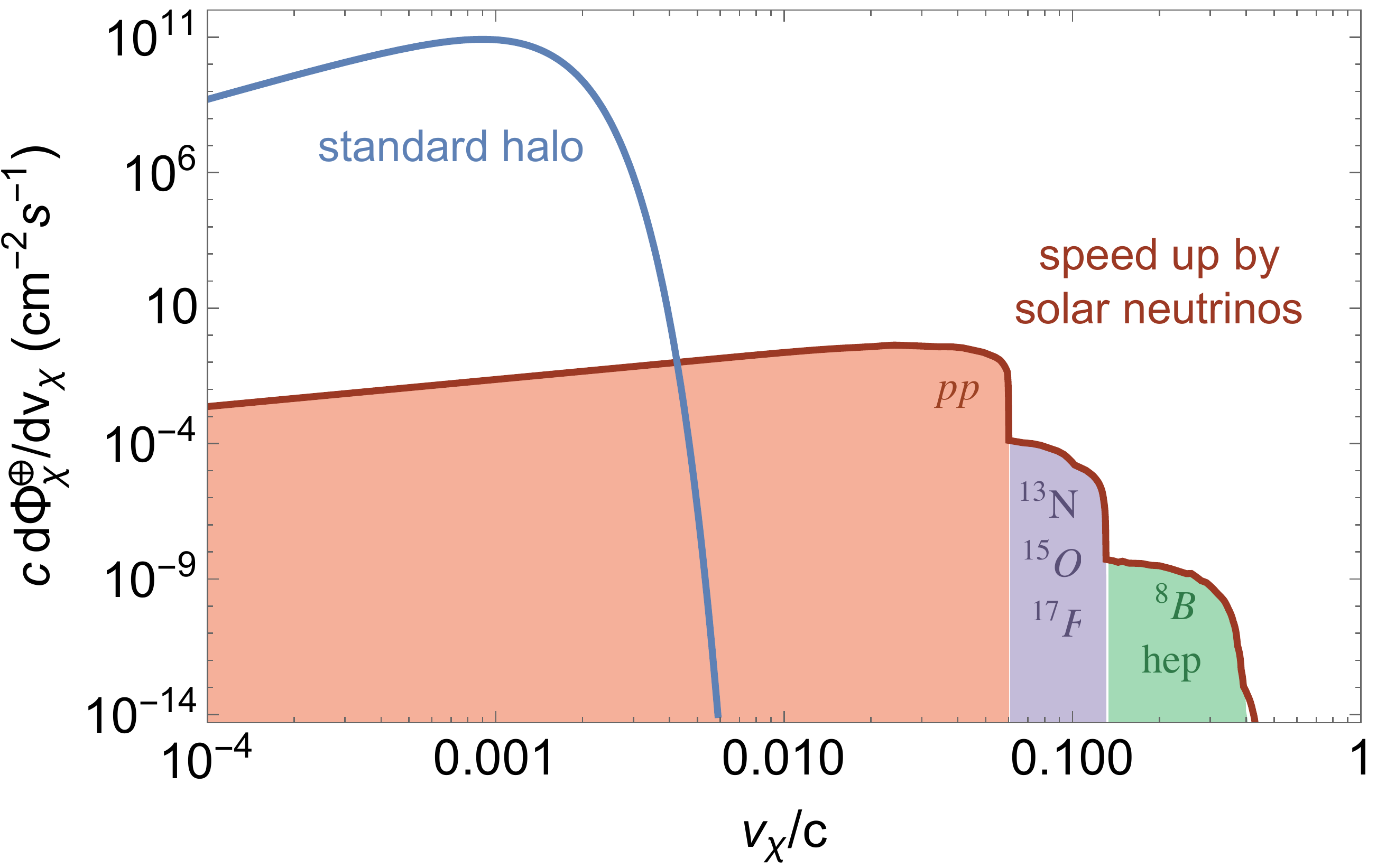}}
\caption{Illustration of dark matter velocity distributions due to the acceleration mechanism discussed in this work (red), decomposed into various solar neutrino components, with parameters $\alpha_D = 0.05$, $v/\Lambda = 0.7$, $m_\chi=0.1\,$GeV, compared with the standard halo velocity distribution (black).}
\label{fig:flux}
\end{figure}

The flux of solar neutrinos depends on the distance from the Sun. It could be calculated based on the measured solar neutrino flux by terrestrial experiments on Earth.
In the reaction $\nu+ B\to \chi + \eta$ we consider, the angular distribution of the final state dark matter particle is isotropic.
Therefore, we integrate over all the space around the Sun for this reaction to occur, and the resulting flux of final state dark matter at Earth is
\begin{equation}\label{eq:Flux}
\frac{d\Phi^\oplus_\chi(E_\nu)}{d E_\nu} = \int d^3 \vec{r} \left[ \frac{\rho_\chi(\vec{r})/(2 m_\chi)}{4\pi d(\vec{r})^2} \right] \left[ \frac{d_\oplus^2}{r^2} \frac{d\Phi^\oplus_\nu(E_\nu)}{d E_\nu} \right] \sigma(E_\nu) \ ,
\end{equation}
where $d(\vec{r})=\sqrt{r^2+d_\oplus^2-2r d_\oplus \cos\theta}$ is the distance from $\vec{r}$ to the Earth's position.
In this coordinate basis, the Sun is located at the origin, the Earth is on the $\hat z$ axis at $d_\oplus=1\,$AU.
$\sigma(E_\nu)$ is the cross section calculated from Eq.~(\ref{eq:crosssectionCoulomb}) and 
${d\Phi^\oplus_\nu(E_\nu)}/{d E_\nu}$ is the differential solar neutrino flux (see {\it e.g.} \cite{Bellerive:2003rj}. We omit the lines).

The angular part of the above integral can be done analytically. The remaining $r$ integral is dominated by the region $r\lesssim d_\oplus$.
Therefore, we can simply set $\rho_\chi(\vec{r}) \simeq \rho_\oplus = 0.3\,{\rm GeV}/{\rm cm}^3$. Under this approximation,
the $r$ integral can also be done analytically, yielding
\begin{equation}\label{eq:FluxFormula}
\frac{d\Phi^\oplus_\chi}{d E_\nu} = \frac{\pi^2 \rho_\oplus  d_\oplus}{8 m_\chi} \sigma(E_\nu) \frac{d\Phi^\oplus_\nu(E_\nu)}{d E_\nu} \ .
\end{equation}
In the dissociation process we consider, the incoming solar neutrino energy is related to the final dark matter velocity $v_\chi$ via Eq.~(\ref{eq:vafter}), $E_\nu \simeq m_\chi v_\chi^2 + E_b$.
We derive a differential dark matter flux with respect to its velocity, 
\begin{equation}\label{eq:DiffernetialFlux}
\frac{d\Phi^\oplus_\chi}{d v_\chi}  = \frac{\pi^2 \rho_\oplus  d_\oplus}{8 m_\chi} \sigma(E_\nu) \left[ 2 m_\chi v_\chi \frac{d\Phi^\oplus_\nu(E_\nu)}{d E_\nu} \right] \ ,
\end{equation}
which serves as new component of dark matter velocity distribution on top of the halo one.
In Fig.~\ref{fig:flux}, we plot this differential flux (red curve) for a set of model parameters, $\alpha_D=0.05$ and $m_\chi=0.1\,$GeV,
and compare it with the dark matter velocity in the standard halo model (black curve).
Clearly, this new component of dark matter particles is characterized by much higher velocities than the halo counterpart. 
In spite of their overall lower flux, these particles could leave a detectable signal above the threshold of many traditional dark matter detectors as will be shown below.

Moreover, the differential flux with respect to dark matter velocity, shown in Fig.~\ref{fig:flux}, looks very similar in shape to its source, the solar neutrino energy spectrum.
Once a positive signal is triggered in direct detection experiments, such a distinct feature may allow us to verify the above dark matter acceleration mechanism.

\section{Reinterpreting Recent PICO Result}
As an application of the accelerated dark matter flux derived above, in this section we explore its implication for dark matter direct detection. 
Here we assume that the dark matter particle $\chi$ has a spin-dependent scattering cross section with the proton (but not with neutron).
We take a phenomenological approach and simply parametrize the cross section as $\sigma_{\chi p}^{\rm SD}$. Following the standard formalism~\cite{Jungman:1995df,An:2010kc}, the corresponding detection rate differentiated with respect to the nuclear recoil energy is
\begin{equation}
\frac{dR}{d E_r} = N_T \frac{m_A }{2 \mu_{\chi A}^2} \int_{v_{min}}^{v_{max}} \frac{dv}{v^2} \frac{d\Phi_\chi^\oplus(v)}{d v} \frac{S(\sqrt{2 m_A E_r})}{S(0)} \sigma_{\chi A}^{\rm SD} \ ,
\end{equation}
where $m_A$ is the mass of the target nucleus, $\mu_{\chi A} = m_\chi m_A/(m_\chi + m_A)$, and $N_T$ is the total number of the nucleus $A$ in the detector. $\sigma_{\chi A}^{\rm SD}$ is the nucleus level cross section, which in the proton-coupling-only case, is equal to $(\mu_{\chi A}^2/\mu_{\chi p}^2) \sigma_{\chi p}^{\rm SD}$. The nuclear form factor is taken from~\cite{Belanger:2008sj}. 
In the standard case, the differential flux is given by $d\Phi_\chi^\oplus/dv = (\rho_\oplus /m_\chi) v f_1(v)$ where $f_1(v)$ is the standard Maxwellian halo velocity distribution~\cite{Jungman:1995df}. 
In contrast, when considering the detection of dark matter speeded up by solar neutrinos, the differential flux is given by Eq.~(\ref{eq:FluxFormula}), with $v_{min}=\sqrt{m_A E_r}/(2\mu_{\chi A}^2)$, and $v_{max}$ corresponding to the highest dark matter velocity after the solar neutrino acceleration (see Fig.~\ref{fig:FD}).

\begin{figure}[t]
\centerline{\includegraphics[width=0.618\textwidth]{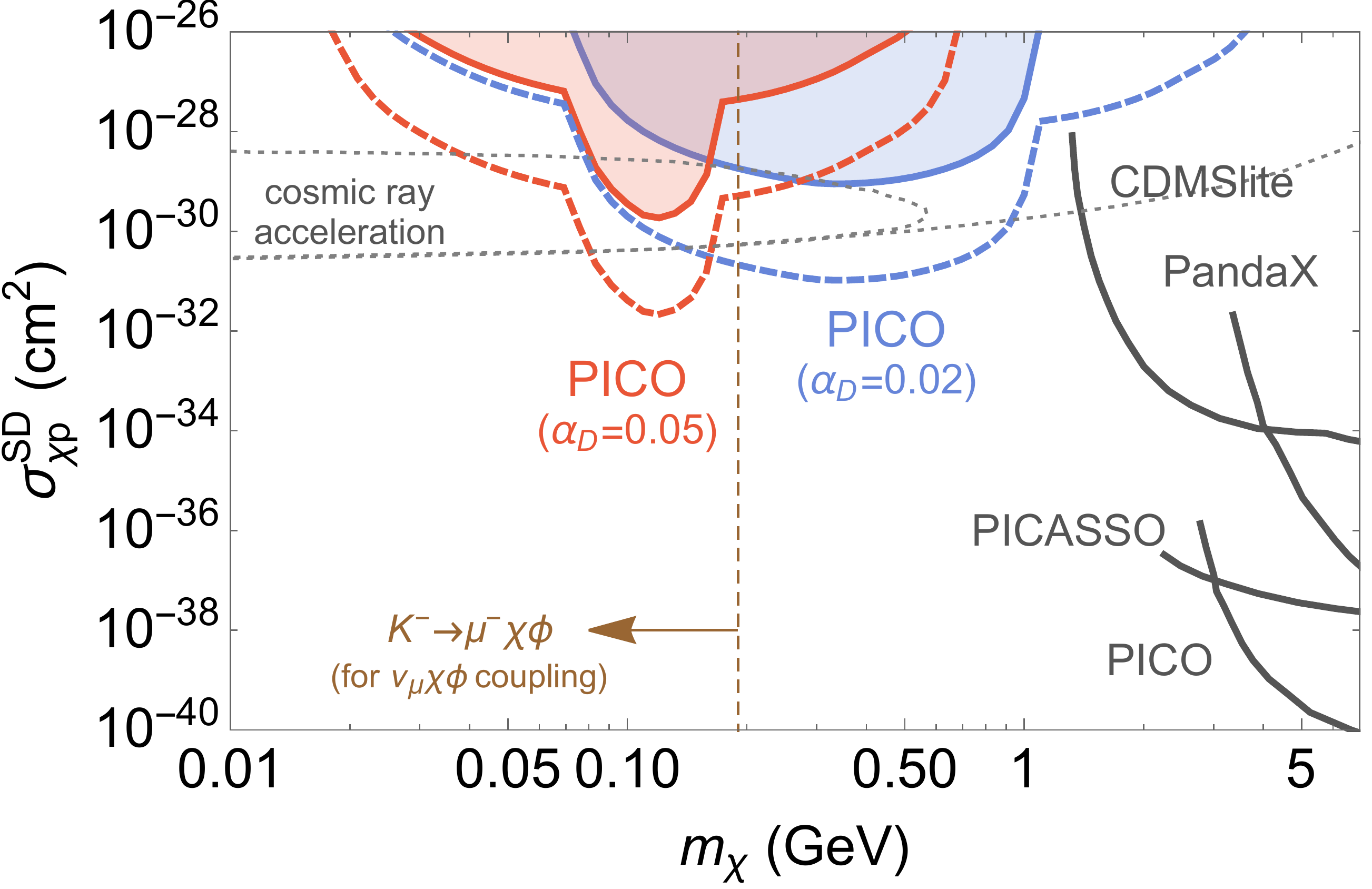}}
\caption{A reinterpretation of the existing (solid) and future (dashed) PICO search results for sub-GeV dark matter, assuming the latter is accelerated by solar neutrinos via the mechanism presented here. 
We fix $v/\Lambda=0.7$ and show results with two benchmark values of $\alpha_D = 0.02$ (blue) and 0.05 (red), respectively.
Upper bound on the cross section from standard direct search (CDMSlite, PandaX, PICASSO, PICO) are shown by the dark gray curves.
Constraint from accelerated dark matter search by cosmic rays excludes the region enclosed by the dotted gray curves~\cite{Bringmann:2018cvk}.
Constraint from charged kaon leptonic decay excludes the region to the left of the brown curve (assuming $m_\chi\simeq m_\phi$), if the neutrino portal interaction involves $\nu_\mu$ flavor.
More details are provided in the main text.
}
\label{fig:reach}
\end{figure}

We consider a recent result from the PICO-60 experiment~\cite{Amole:2017dex} at SNOLAB, which is based on a 1167 kg-day exposure using C$_3$F$_8$ and reinterpret the result for the speeding-up scenario. 
The result is presented in the $\sigma_{\chi p}^{\rm SD}$ versus $m_\chi$ plane in Fig.~\ref{fig:reach}.
The red and blue shaded regions are the new exclusion limits that the PICO-60 experiment could set for if dark matter particles are accelerated by solar neutrinos as considered in this work, for the dark fine-structure constant $\alpha_D=0.05$ and $0.02$, respectively. 
For comparison, we also include results from PICO60~\cite{Amole:2017dex}, PICASSO~\cite{Behnke:2016lsk}, CDMSlite~\cite{Agnese:2017jvy} and PandaX~\cite{Xia:2018qgs}, assuming standard halo model.

For the above limits to apply, the dark matter particles need to successfully travel to the underground detector, instead of shielded by the earth above.
For spin-dependent scattering, the main target is the $^{27}$Al in the Earth's crust. Using the crust mass density $2.6\,{\rm g/cm^3}$ and the mass fraction of $^{27}$Al 8\%~\cite{chemical}, we derive the number density of $^{27}$Al nucleus $n_{\rm Al} \sim 10^{21}\,{\rm cm^{-3}}$. The mean free path of dark matter in the earth's crust is given by $1/(n_{\rm Al} \sigma)$. For this length to be larger than the depth of PICO detector ($\sim 2\,$km), we need $\sigma \lesssim 10^{-26}\,{\rm cm}^2$.
This sets the upper range for the cross section in Fig.~\ref{fig:reach}.

The shape of the new exclusion region derived in this work can be understood through the following considerations. Two conditions must be satisfied for the scenario we consider to occur. First, the solar neutrino must be energetic enough to dissociate the dark matter bound state, which requires 
\begin{equation}\label{lowerboundvchi}
E_\nu > E_b \simeq \alpha_D^2 m_\chi/4 \ ,
\end{equation}
in the assumed Coulomb limit. 
Second, the recoil energy from the resulting dark matter scattering must exceed the experimental threshold, 
\begin{equation}
m_\chi E_\nu/(2 m_A) > E_{\rm th} \ .
\end{equation}
For PICO-60, $E_{\rm th}=3.3\,$keV. For a particular solar neutrino component ({\it e.g.} $pp$ neutrinos with energies $E_\nu\lesssim 0.4\,$MeV) and the value of $\alpha_D$, there is a window of $m_\chi$ (corresponding to the bumps in the exclusion region in Fig.~\ref{fig:reach}) to satisfy both conditions, $2 m_A E_{\rm th}/E_\nu < m_\chi < 4 E_\nu/\alpha_D^2$. 
Therefore, if the dark matter mass is too large or too small, the limit weakens quickly. 
Moreover, the condition for the above window to exist is $\alpha_D < \sqrt{2 E_\nu^2/ (m_A E_{\rm th})}$.
As $\alpha_D$ grows, such a window narrows and finally disappears. In turn, one has to resort to higher energy solar neutrino components at the price of lower fluxes. In Fig.~\ref{fig:reach}, the dashed curves show the potential future reach of PICO experiment assuming a $10^5\,$kg-day exposure~\cite{Levine:2018talk} and the same energy threshold. 
The reaches can be further improved with a lower detector threshold~\cite{Amole:2019scf}.

\section{Other Constraints}
Here we discuss other constraints on the model we consider.
First, the effective operator introduced in Eq.~\eqref{EFO} allows for new invisible decay channels of the Higgs and $Z$ bosons, $h, Z\to\nu+\chi+\phi$. 
The latest constraint on $Z$ and Higgs invisible widths set an upper bound $v/\Lambda \lesssim0.7$ for $\chi$ and $\phi$ mass in the range of a few hundred MeV~\cite{Orlofsky:2021mmy}.
In the above analysis, we have chosen $v/\Lambda=0.7$ to be consistent with this limit.
Second, for sufficiently light $\chi$ and $\phi$ particles, the neutrino portal interaction can also lead to $K^-\to \mu^- + \chi + \phi$ decay, if the neutrino is of the $\nu_\mu$ flavor~\cite{Batell:2017cmf}. 
The signature is a distortion of the final state muon spectrum which would be mono-energetic in the rest frame of kaon from regular leptonic decay.
With $v/\Lambda \sim \mathcal{O}(1)$, this constraint excludes the mass range $m_\chi+m_\phi < m_K - m_\mu \simeq 0.39\,$GeV.
It corresponds to the region to the left of the vertical line in Fig.~\ref{fig:reach}.
Third, the dark-matter-nucleon interaction could also be constrained by looking for cosmic-ray accelerated dark matter particles (see e.g. \cite{Bringmann:2018cvk}).
The corresponding limits are shown by the gray dotted curves in Fig.~\ref{fig:reach}. This constraint is subjected to uncertainties in the cosmic ray distribution in our galaxy.
Finally, we also note that the neutrino portal interaction could also induce dark-matter-electron via additional weak interaction insertion.
Detailed exploration of the potential signals is beyond the scope of this work.

\section{Conclusion}
In this work, we explore the possible role of solar neutrinos on the dark matter velocity distribution and direct detection experiments.
In a simple neutrino portal dark sector model, we show it is possible to speed up dark matter well above $10^{-3}c$. 
A crucial ingredient for this mechanism to work is the existence of stable bound states of dark matter.
The accelerated dark matter particles feature characteristic velocity distribution which could make this mechanism testable.
When such a particle strikes on the nucleus target in a dark matter detector, the recoil energy is large enough to render useful limits on sub-GeV dark matter interaction, by 
reinterpreting the existing direct detection results.
Although we only did so based on a spin-dependent search result from PICO-60, it is straightforward to generalize this idea to other dark matter direct detection results.
The parameter space of interest to this study can also be probed in terms of Higgs, $Z$-boson and meson decays at the precision frontiers.
Last but not least, the model considered here features a light dark force carrier $\phi$ which if sufficiently light may leave an imprint on the big-bang nucleosynthesis. 

Given that the nature of dark matter remains unknown and is being pursued by an intensive program of direct searches,
it is important to keep an open mind to novel acceleration mechanisms that may reveal light dark matter as discussed in this work.

\section*{Acknowledgment}

I would like to thank Eric Dahl for a useful discussion on future spin-dependent dark matter detection experiments. This work is supported by the Arthur B. McDonald Canadian Astroparticle Physics Research Institute.

\bibliographystyle{unsrt}
\bibliography{references.bib}

\end{document}